\documentclass[twocolumn,showpacs,preprintnumbers,amsmath,amssymb,aps]{revtex4}
\usepackage{dcolumn}
\usepackage{bm}
\usepackage{graphicx}
\usepackage{amsmath}
\DeclareMathOperator{\Tr}{Tr}
\usepackage{changes}

\begin{document}

\title{Selective nuclear-spin interaction based on a dissipatively stabilized nitrogen-vacancy center}
\author{Jiawen Jiang and Q. Chen\footnote{E-mail:qchen@hunnu.edu.cn}}
\affiliation{ Department of Physics and Key Laboratory of Low
Dimensional Quantum Structures and Quantum
Control of Ministry of Education, Hunan Normal University, Changsha 410081, China.\\
}

\begin{abstract}
Current typical methods to realize nuclear-nuclear quantum gates require a sequence of electron-nuclear quantum gates by using dynamical decoupling techniques, which are implemented at low temperature because of short decoherence and relaxation time of the NV spin at room temperature. This limitation could be overcome by using periodical resets of an NV spin as a mediator of interaction between two nuclear spins  [Chen, Schwarz, and Plenio, 119, 010801 (2017)]. However, this method works under stringent coupling strengths condition, which makes it not applicable to heteronuclear quantum gate operations. Here we develop this scheme by using radio-frequency (RF) fields to control different nuclear spin species. Periodical resets of the NV center protect the nuclear spins from decoherence and relaxation of the NV spin. RF control provides probability to have highly selective and high fidelity quantum gates between heteronuclear spins as well as detecting nuclear spins by using a nuclear spin sensor  under ambient conditions.
\end{abstract}

\maketitle

\section{Introduction}
Nuclear spins in material such as diamond associated with single defects represent a promising platform for quantum information registers \cite{Taminiau2014UniversalCA,2018Pulse,Waldherr2014QuantumEC,2011High} and sensing purposes \cite{Taminiau2012DetectionAC,Kolkowitz2012SensingDN,Zhao2012SensingSR,Ermakova2013DetectionOA,Mamin2013NanoscaleNM,Zaiser2016EnhancingQS,
Staudacher2013NuclearMR,Wu2016DiamondQD} due to their long coherence times and the potentially large number of available spins. Nuclear spins can be initialized, controlled, and read out through the electron spin of the nitrogen vacancy (NV) center \cite{Doherty2013TheNC} driven by optical fields and microwave radiation \cite{2016Optomechanical,2014Optically}.  Recent progress manifests in a number of works, such as electron-nuclear \cite{Casanova2016NoiseResilientQC,Bermdez2011ElectronmediatedNI,zimmermann2020selective,rong2014implementation,hegde2020efficient,
abobeih2019atomic,Bradley2019ATS,Wang2017DelayedEE,Tratzmiller2021ParallelSN,Casanova2017ArbitraryNG}, electron-electron \cite{degen2021entanglement} and nuclear-nuclear quantum gates \cite{Bian2017UniversalQC}. Many schemes are proposed to realize electron-nuclear quantum gate operations, both with \cite{Bradley2019ATS,Wang2017DelayedEE} and without \cite{Tratzmiller2021ParallelSN,Casanova2017ArbitraryNG} additional radio-frequency (RF) control on the nuclear spins themselves.

Nuclear-nuclear quantum gate is implemented by a sequence of the NV-nuclear quantum operations \cite{Taminiau2014UniversalCA,2018Pulse}, and it is a delicate issue to have a complete set of quantum gates on specific nuclei in samples \cite{Bradley2019ATS}.  Current related experiments are operated at low temperature, because of the relatively short lifetime of the NV center. Several outstanding challenges are caused by the relaxation and decoherence processes of the electron spin as these limit quantum gate fidelities on nuclear registers as well as spectral resolution and selectivity. A scheme was proposed to have highly selective and high fidelity quantum gates between nuclear spins under ambient conditions \cite{Chen2017DissipativelySQ}. However, it has strict requirements of resonance conditions and is applicable for special cases, i.e., for nuclear spins in the same species with the parallel coupling components between the NV and nuclear spins far smaller than the vertical components.


Here we extend the scheme in Ref. \cite{Chen2017DissipativelySQ} to a heteronuclear case by using radio-frequency fields to control different nuclear spin species individually. The effective substantial second-order coupling between the nuclear spins obtained through a MW driving NV center which is periodically reinitialized by a dissipative process \cite{Chen2017DissipativelySQ}. Similarly, the periodical reinitialization of the NV center decouple it from the dynamics and its effect on system is an effective weak dissipation process. Thus high selectivity and fidelity of nuclear-spin quantum gate for different nuclear spin species could be possible even at ambient condition, which is an extension for quantum computation and simulation applications by using different nuclear species controlled by NV centers. Additionally, one can use a nuclear spin as a quantum sensor to detect nuclear spins in another species as well as analysis of complex spin structures. 

This paper is structured as follows. We start with the derivation of effective Hamiltonian of the system by using Schrieffer-Wolf transformation\cite{Kessler2012GeneralizedSF,Bravyi2011SchriefferWolffTF}, which demonstrates the validity of the indirect interaction between nuclear spins by the application of suitably tuned RF fields. We take two relevant species, i.e., carbon-13 and silicon-29 spins as examples to show our protocol efficiency. Coherent evolution between a carbon-13 and a silicon-29 spin is limited by the NV lifetime, this limitation is overcome through periodical resets of NV spin. The feasibility of high selectivity and fidelity nuclear-nuclear quantum gates by using RF controls is discussed as well as the sensing application. Finally, we compare our approach to the previous scheme in Ref. \cite{Chen2017DissipativelySQ}.

\section{Indirect interaction between nuclear spins}

\begin{figure*}[t]
\center
\includegraphics[width=6.8in]{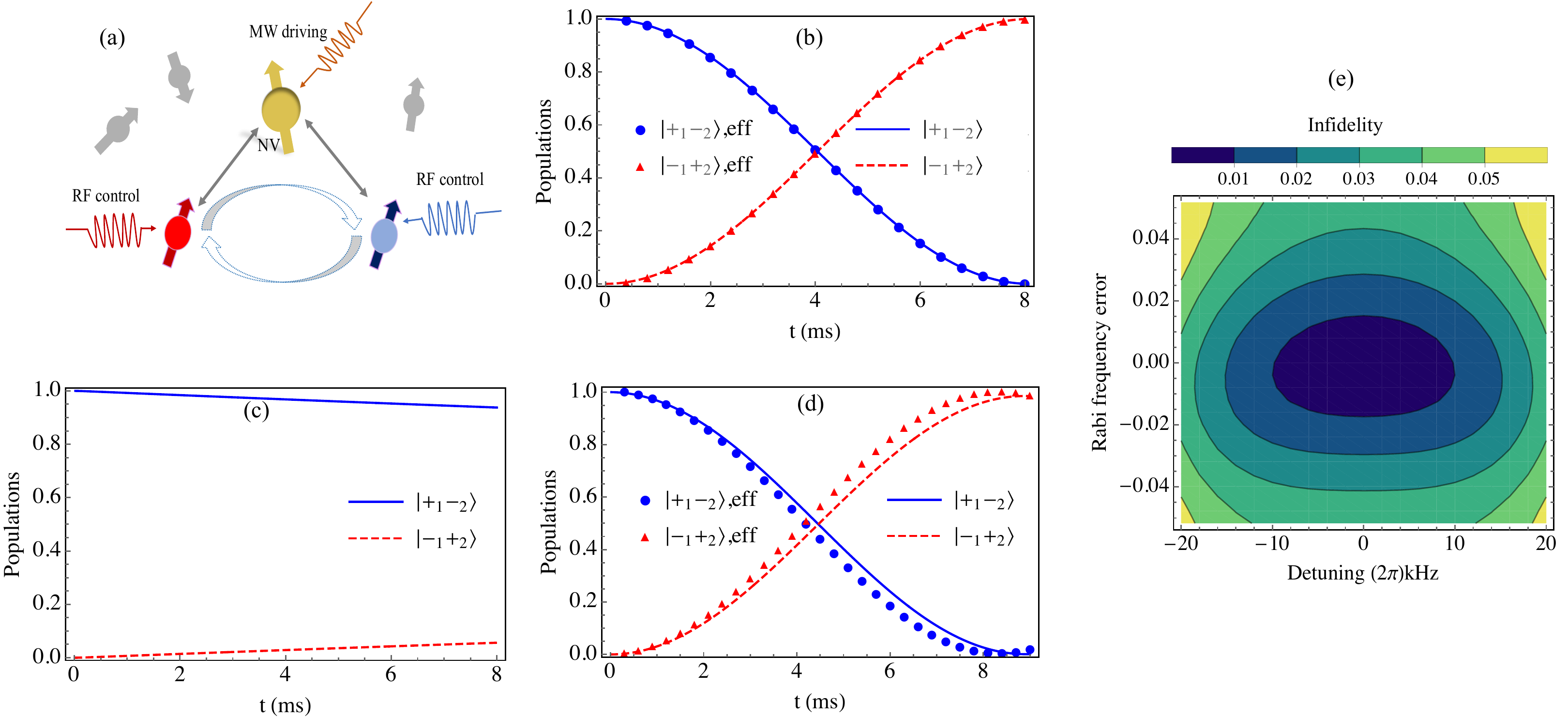}
\caption{(a) The NV center mediates the coupling between a silicon-29 spin with  $\gamma_{n_1}B=(2\pi)4$ MHz and a carbon-13 spin $\gamma_{n_2}B=(2\pi)5.06$ MHz to achieve a quantum gate, while itself is decoupled from the dynamics. RF fields are applied to control two nuclear spin species individually, when we use continual MW field to drive the NV spin. (b) Population evolutions are shown by considering the ideal case in which no dissipation is included and the nuclear spins is initialized in state $|+_{1}-_{2}\rangle$. The Rabi frequencies of RF fields are given by $\Omega_{rf1}=\Omega_{rf2}=(2\pi) 1$ kHz, microwave driving field $\Omega=(2\pi) 400$ kHz, the parallel coupling components $[a_{\parallel 1},a_{\parallel 2}]=(2\pi) [9,11]$ kHz and the effective coupling coefficient $g_{e}=(2\pi) 0.12$ kHz. These exact numerical simulations by using Eq. (\ref{H't}) fit well to the theoretical derivation of the effective (eff) dynamics under Hamiltonian Eq. (\ref{He}). (c) By considering the practical case, we use the same parameters in (b) and the NV dissipation is included as $T_{1\rho}=200$ $\mu s$, exact numerical simulation of the master equation Eq. (\ref{master}) shows that there is no coherent evolution between the nuclear spins due to the NV life time limitation.  (d) The periodical NV resets makes it possible to extend the coherent evolution of the heteronuclear spins well beyond the NV spin lifetime. The exact numerical calculation is based on Eq. (\ref{Evolution}), when we simulate the effective master equation Eq. (\ref{masterN}) of nuclear spins. The parameters are as the same as in (b) except that the NV center is reinitialied to state $|-\rangle_e$ every $t_{re}=20$ $\mu s$. (e) Process infidelity for different detuning and errors of the MW driving of the NV center. }
\label{NoReset}
\end{figure*}

We consider two nuclear spins in different species are coupled to an electron spin of single NV center, and microwave (MW) field and radio-frequency (RF) field are used for external control over the electron and nuclear spins as well as for achieving selective internuclear interactions, see Fig. \ref{NoReset}(a). The magnetic field $B$ is applied along the NV axis (the $\hat{z}$ axis), which is large enough to split the degenerate states of $|m_s=|\pm1\rangle$. The Hamiltonian of the whole system is given by
\begin{eqnarray}
H&=&DS_{z}^{2}+\gamma_{e}BS_z+\sum_{i}\gamma_{n_i}BI_i^z \\
& &+S_z\sum_{i}\overrightarrow{A_i}\cdot\overrightarrow{I_i}+H_{w}+H_{r}. \notag
\label{eq1}
\end{eqnarray}
Note that we hereafter omit the Dirac constant $\hbar$ for simplicity. Here $D=(2\pi)2.87$ GHz denotes zero-field splitting of the electronic ground state, $\gamma_{e}$ and $\gamma_{n_i}$ are the gyromagenetic ratio of the electron spin and nuclear spins, respectively. The interaction between the NV center
and the $i$th nucleus is mediated by the hyperfine vector $\overrightarrow{A_i}$, $A_{i}=(a_{\parallel i},a_{\perp i})$ with $a_{\parallel i}$ and $a_{\perp i}$ denotes the related coupling parallel and perpendicular components to the nuclear spin quantization axes $a_{\parallel i}=\overrightarrow{A}_{i}\cdot\hat{z}$ and $a_{\perp i}=\sqrt{|\overrightarrow{A}_{i}|^{2}-a^{2}_{\parallel i}}$. Hamiltonians $H_{w}$ and $H_{r}$ describe the action of the MW driving and RF fields, respectively, which reads



%
\begin{eqnarray}
H_{w}&=&\sqrt{2}\Omega\cos\omega tS_{x}, \\
H_{r}&=&\sum_{i}2\Omega_{rfi}\cos\omega_{rfi}tI_{i}^{x},
\label{driving}
\end{eqnarray}
where $\Omega$ ($\Omega_{rfi}$) is the Rabi frequency of the MW (RF) driving field with the corresponding frequency $\omega$ ($\omega_{rfi}$).

The MW driving field is applied to be on resonance of the transition $|-1\rangle\leftrightarrow|0\rangle$ as $\omega=D-\gamma_eB$, we can rewrite the effective Hamiltonian of the NV center as $H_{NV}=\Omega\sigma_{z}$.  Then in the rotating frame with $H_0=(D-\gamma_eB)|-1\rangle\langle-1|$, the Hamiltonian could be rewritten as
\begin{eqnarray}
H_{t}&=&\Omega\sigma_{z}+\sum_{i}(\gamma_{ni}B+\frac{a_{\parallel i}}{2})I^{z}_{i} \notag\\
& &+a_{\parallel i}\sigma_{x}I^{z}_{i}+a_{\perp i}\sigma_{x}I^{x}_{i}+H_{r},
\label{Ht}
\end{eqnarray}
where $\sigma_{z}=\frac{1}{2}(|+\rangle_e\langle+|-|-\rangle_{e}\langle-|)$ with $|+\rangle_{e}=\frac{1}{\sqrt{2}}(|0\rangle+|-1\rangle)$ and $|-\rangle_{e}=\frac{1}{\sqrt{2}}(|0\rangle-|-1\rangle)$.

Two weak RF fields are applied to individually control two different nuclear spins with $\omega_{rfi}=\gamma_{n_i}B+\frac{a_{\parallel i}}{2}$. Working in the rotating frame with $H'_{0}=\sum_i(\gamma_{ni}B+\frac{a_{\parallel i}}{2})I^{z}_{i}$, by assuming $a_{\parallel i},a_{\perp i}\ll\Omega\ll\omega_{rfi}$ and $a_{\parallel i}$ is comparable or larger than $a_{\perp i}$, and the total Hamiltonian of the system can be approximated as 
\begin{eqnarray}\label{eq3}
H'_{t}&=&\Omega\sigma_{z}+\sum_{i}\Omega_{rfi}I^{x}_{i}+a_{\parallel i}\sigma_{x}I^{z}_{i}.
\label{H't}
\end{eqnarray}
%
Consider the Hamiltonian is of the form $H'_{t}= H'_{NV}+V$, where $H'_{NV}=\Omega\sigma_{z}$ denotes the Hamiltonian of the MW driving and $V=\sum_{i}\Omega_{rfi}I^{x}_{i}+a_{\parallel i}\sigma_{x}I^{z}_{i}$ is the weak perturbation with $a_{\parallel i},\Omega_{rfi}\ll\Omega$. By using the Schrieffer-Wolff transformation \cite{Kessler2012GeneralizedSF,Bravyi2011SchriefferWolffTF} in condensed matter, the second order expansion due to perturbation terms $V$ can be obtained as
 \begin{eqnarray}
      \langle\alpha|H_{e}|\beta\rangle_e=\frac{1}{2}\sum_{i}(\frac{\langle\alpha|V|i\rangle\langle i|V|\beta\rangle_e}{E_\alpha-E_i}-\frac{\langle\alpha|V|i\rangle\langle i|V|\beta\rangle_e}{E_i-E_{\beta}}),
    \label{SW}
\end{eqnarray}
in which $\alpha,\beta,i=\{+,-\}$, $H_e$ is defined as the effective Hamiltonian of the system and $E_k$ ($k=\{+,-\}$) is the energy corresponding to the state $|k\rangle_e$. By simple calculations, we find $\langle+|H_{e}|+\rangle_e=g_{e}I^{z}_{1}I^{z}_{2}$ in which $|i\rangle_e=|-\rangle_e$ is the channel state of virtual electron spin flip, when $\langle-|H_{e}|-\rangle_e=- \langle+|H_{e}|+\rangle_e=g_{e}I^{z}_{1}I^{z}_{2}$ with the channel state $|i\rangle_e=|+\rangle_e$, here $g_{e}\approx\frac{a_{\parallel 1}a_{\parallel 2}}{2\Omega}$. Thus we adiabatically eliminate the fast electronic degrees of freedom from the slow nuclear dynamics by a Schrieffer-Wolff (SW) transformation, and derive the effective Hamiltonian
%
\begin{eqnarray}\label{He}
H_{e}\approx\sum_{i=1,2}\Omega_{rfi}I^{x}_{i}+g_{e}I^{z}_{1}I^{z}_{2}\otimes(|+\rangle_e\langle+|-|-\rangle_e\langle-|).
\end{eqnarray}
To investigate the dynamics of the whole system described by density matrix $\rho$, the system is governed by the master equation
\begin{eqnarray}
\label{masterT}
\frac{d\rho}{dt}=-i[H'_{t},\rho]+ L_e\rho L_e^{\dagger}-\frac{1}{2}(L_eL_e^{\dagger} \rho+\rho L_eL_e^{\dagger})
\end{eqnarray}
in which $L_e=\sqrt{\gamma_e}|-\rangle_e\langle+|$ with  $\gamma_{e}=1/T_{1\rho}$. We apply this method and evaluate the performance of the gate between the first spin (carbon-13) with  $\gamma_{n_1}B=(2\pi)4$ MHz and the second spin (silicon-29) $\gamma_{n_2}B=(2\pi)5.06$ MHz coupled to an NV center with $[a_{\parallel 1},a_{\parallel 2}]=(2\pi)[9,11]$ kHz, [see Fig. \ref{NoReset}(a)]. Here the driving fields are given as RF Rabi frequencies $\Omega_{rf1}=\Omega_{rf2}=(2\pi)1$ kHz, and the MW Rabi frequency $\Omega=(2\pi)400$ kHz. The NV center is initialized to state $|-\rangle_{e}$, when the nuclear spins are initialized to $|+_{1}-_{2}\rangle$ with $|+_{i}\rangle=\frac{1}{2}(|\uparrow_{i}\rangle+|\downarrow_{i}\rangle)$. Perfect state transfer between nuclear spins silicon-29 and carbon-13 $|+_{1}-_{2}\rangle\rightarrow|-_{1}+_{2}\rangle$ is shown in Fig. \ref{NoReset}(b), in which the population evolutions by using effective second order Hamiltonian Eq. (\ref{He}) fit well with the exact total Hamiltonian of the system Eq. (\ref{H't}). The effective Hamiltonian Eq. (\ref{He}) includes a three-body interaction and there are two channels that can mediate internuclear interaction via virtual electron spin flips through the NV microwave dressed states $|+\rangle_{e}$ and $|-\rangle_{e}$. However, the two channels can be mixed up if the nuclear-nuclear interaction time is longer than the NV relaxation time. As shown in Fig. \ref{NoReset}(c), in the case when the life time of NV center $T_{1\rho}=200$ $\mu s$, there is no coherent evolution of the nuclear spins. Therefore, the system is limited by the NV relaxation at room temperature.

\begin{figure*}[t]
\center
\includegraphics[width=6.6in]{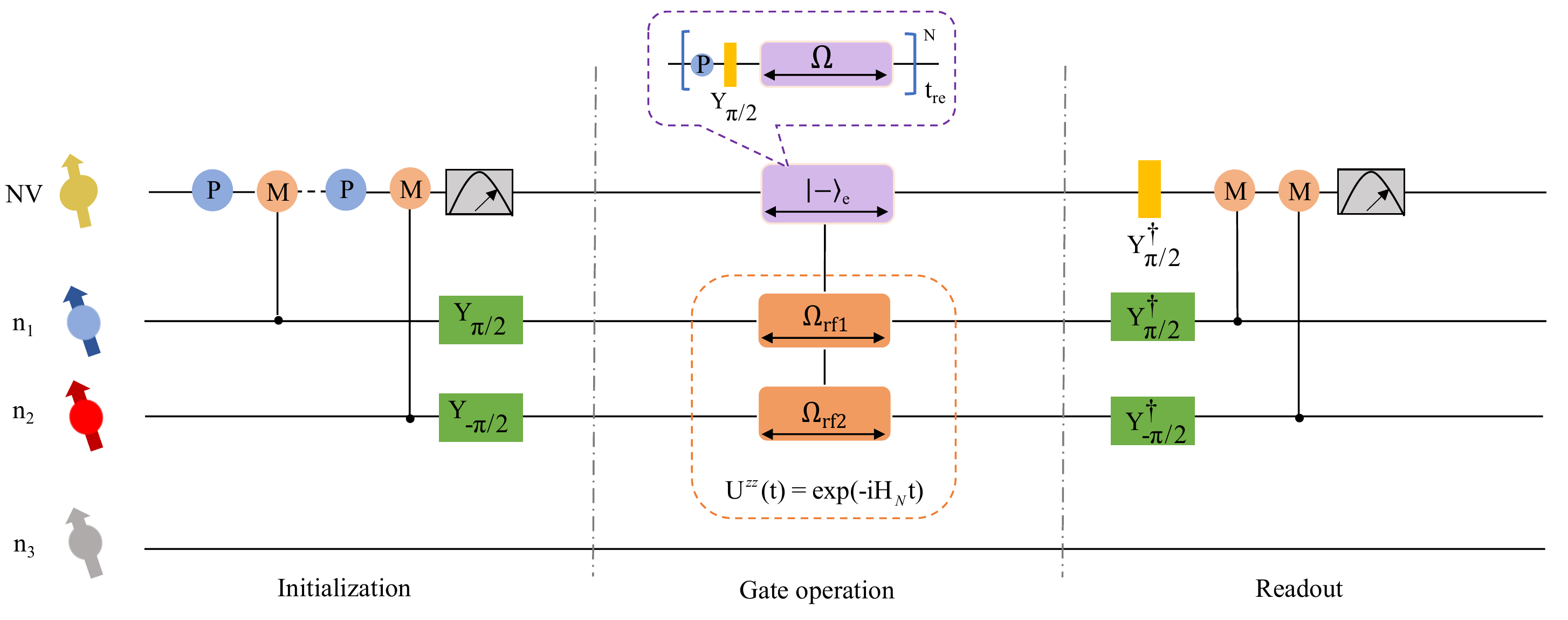}
\caption{Nuclear-nuclear quantum gate of two nuclear spins mediated by an NV center in diamond. The basic logic-operations contain the initialization, gate operation based on second-order coupling and readout of nuclear spins. The electron spin works as a quantum bus and is stabilized in a quasi-steady state by periodically resetting in $|-\rangle_{e}$, which is shown in the yellow dashed box. Here P and M stand for the electron (nuclear) spin polarization, $Y_{\theta}$ presents operation with rotation angles $\theta$ about the $Y$ axes and $N$ is the total number of resets.}
\label{Operation}
\end{figure*}

\section{The effect of electron spin resets}
In our scheme, in order to overcome the limitation of NV relaxation, we intend to periodical reset the electron spin to state $|-\rangle_{e}$ within its relaxation time $T_{1\rho}$. It is reasonable to consider the steady-state polarization of the NV spin in a quasi-steady state as
\begin{eqnarray}
\langle2\sigma_z\rangle=e^{-t_{re}/T_{1\rho}}=p.
\end{eqnarray}
The three-body interaction in the effective Hamiltonian Eq. (\ref{He}) can be estimated as $g_{e}I^{z}_{1}I^{z}_{2}\otimes(|+\rangle_e\langle+|-|-\rangle_e\langle-|)\approx g_{e}I^{z}_{1}I^{z}_{2}\langle2\sigma_z\rangle=pg_{e}I^{z}_{1}I^{z}_{2}$. Therefore, the NV spin is not involved in the nuclear-nuclear interaction Hamiltonian. Resets also bring another dissipation item in the master equation and it is reasonable to have the effective relaxation of the NV spin as $\gamma_{r}=1/T_{1\rho}+1/t_{re}$. Thus, we can rewrite the master equation of the system as
\begin{eqnarray}
\label{master}
\frac{d\rho}{dt}=-i[H'_{t},\rho]+L_r \rho L_r^{\dagger}-\frac{1}{2}(L_r ^{\dagger}L_r \rho+\rho L_r ^{\dagger}L_r)
\end{eqnarray}
in which $L_{r}=\sqrt{\gamma_r}|-\rangle_e\langle+|$. The non-Hermitian Hamiltonian of the quantum jump formalism \cite{Reiter2012EffectiveOF} of the NV spin is given by
\begin{eqnarray}
H_H=\Omega\sigma_z-\frac{1}{2}L_r^{\dagger}L_r
\end{eqnarray}
According to the second expansion of SW transformation, we have energy difference between $|+\rangle_e$ and $|-\rangle_e$ as a complex energy $E_+-E_-=\Omega-\frac{i}{2}\gamma_r$, one can derive effective Hamiltonian
\begin{eqnarray}
H'_{e}\approx\sum_{i}\Omega_{rfi}I^{x}_{i}+g'_{e}I^{z}_{1}I^{z}_{2}\otimes(|+\rangle_e\langle+|-|-\rangle_e\langle-|),\\
L_n\approx\sum_{k}\frac{\sqrt{\gamma_r}a_{\parallel k}}{\Omega-\frac{i}{2}\gamma_r}I_k^z\otimes(|+\rangle_{e}\langle+|-|-\rangle_{e}\langle-|),
\end{eqnarray}
in which the effective dissipation item is given by $\langle \alpha |L_e|\beta\rangle=\sum_{i}\frac{\langle\alpha| L_r|i\rangle\langle i|V|\beta\rangle}{E_\alpha-E_i}$, and the effective coupling is $$g'_e\approx\frac{\Omega a_{\parallel 1}a_{\parallel 2}}{2(\Omega^2+\frac{\gamma_r^2}{4})}.$$ By considering the electron spin is periodically reset in state $|-\rangle_e$ with the definition $\langle 2\sigma_z\rangle=p$, we have the effective master equation of the nuclear spins as
\begin{eqnarray}\label{masterN}
\frac{d}{dt}\rho_n&=&-i[H_{N},\rho_{n}]+L_{N} \rho_n L_{N}^{\dagger} \notag \\
& &-\frac{1}{2}(L_{N} ^{\dagger}L_{N} \rho_n+\rho_n L_{N} ^{\dagger}L_{N}),
\end{eqnarray}
with the effective Hamiltonian and Lindblad operators
\begin{eqnarray}\label{HLN}
H_{N}&\approx&\sum_{i}\Omega_{rfi}I^{x}_{i}+pg'_{e}I^{z}_{1}I^{z}_{2},\\
L_{N}&\approx&\sum_{k}\frac{\sqrt{p\gamma_r}a_{\parallel k}}{\Omega-\frac{i}{2}\gamma_r}I_z^k.
\end{eqnarray}
The effective dissipation rate is given by
$$\gamma_N=\sum_{i,j}\frac{p\gamma_ra_{\parallel i}a_{\parallel j}}{\Omega^2+\frac{\gamma_r^2}{4}}.$$ Therefore, in principle, if resonant condition $\Omega_{rf1}=\Omega_{rf2}$ and $pg'_{e}\gg\gamma_N$ are satisfied, perfect coherent state transition between the nuclear spins can be possible.

\begin{figure}
\center
\includegraphics[width=3.1in]{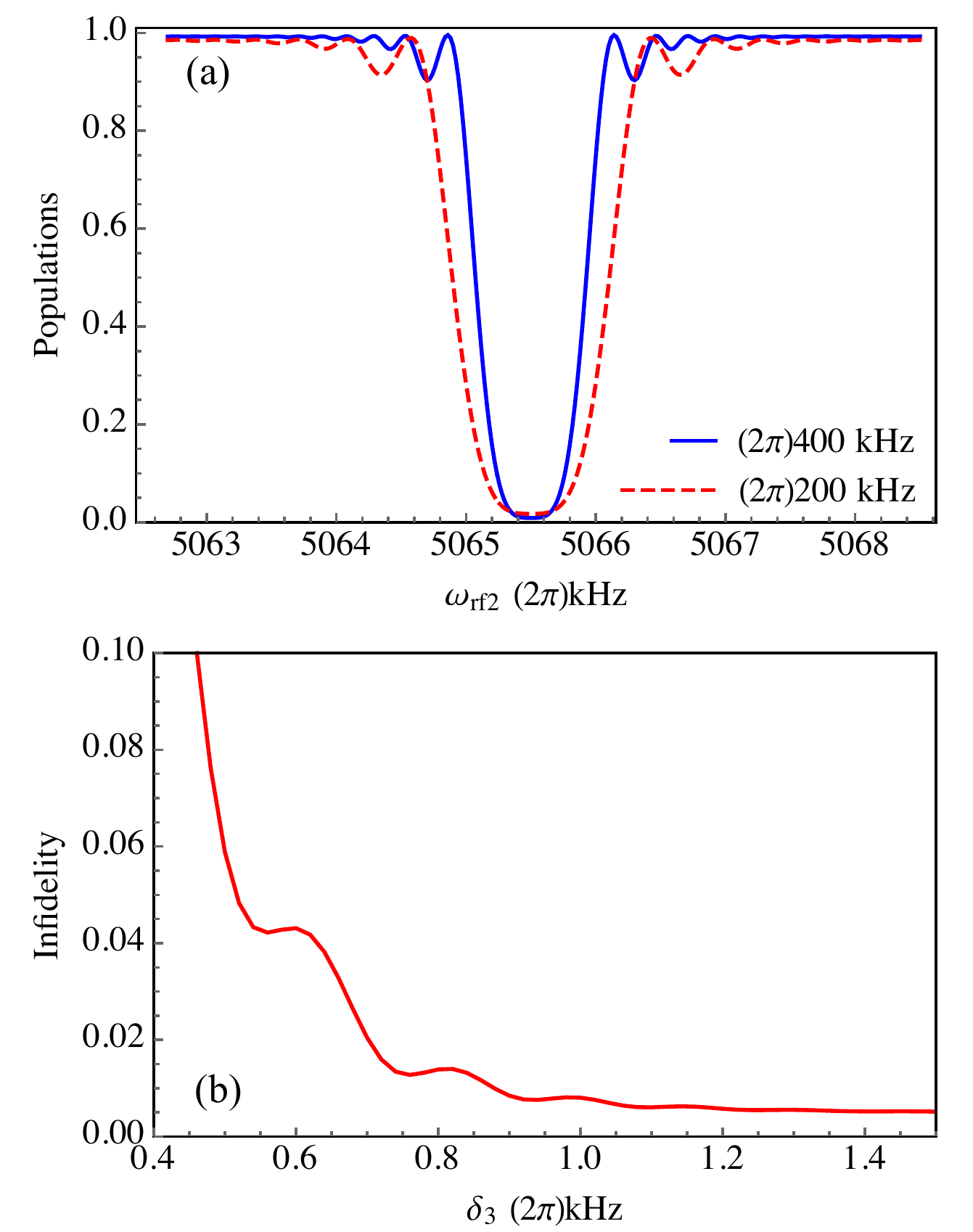}
\caption{Numerical simulations are show with the same parameters in the Fig. \ref{NoReset}, i.e., parallel couplings are given as $[a_{\parallel 1},a_{\parallel 2}]=(2\pi)[9,11]$ kHz. (a) The detection bandwidth of the swept RF frequency based on different MW Rabi frequencies. The RF field is applied to the carbon-13 sensor resonantly when the frequency of the other RF field is swept. The evolution time is fixed as $T=[8.8,4.4]$ ms for $\Omega=(2\pi)[400,200]$ kHz. (b) The infidelity of quantum gate between carbon-13 and silicon-29 spins when there is another silicon-29 spin. The infidelity varies with the detuning $\delta_{3}$, $\delta_{3}=(a_{\parallel 3}-a_{\parallel 2})/2$. }
\label{Robust}
\end{figure}

For comparison, it is necessary to employ exact numerical simulations as well to show the efficiency of our theory. Exact state evolutions of the system are simulated as follows.  The NV center is reinitialized periodically to the state $|-\rangle_e$ of the dressed state basis, namely, $\rho(Nt_{re})\rightarrow[\Tr_{e}\rho(Nt_{re})]\otimes|-\rangle_e\langle-|$, where $\Tr_{e}$ denotes the partial trace over the electron spin and N is an integer and reset of the NV spin every $t_{re}$ introduces an effective interaction time in each cycle. The density matrix of the system evolves according to
\begin{eqnarray}\label{Evolution}
\rho_{n}\rightarrow\cdot\cdot\cdot U_{t}\Tr_{e}[U_{t}(\rho_{n}\otimes|-\rangle_e\langle-|)U^{\dagger}_{t}]\otimes|-\rangle_e\langle-|U^{\dagger}_{t},
\end{eqnarray}
in which $U_{t}$ is the time evolution operator according master equation Eq. (\ref{master}). Both of the state evolutions by using exact numerical simulation and effective second-order master equation based on  Eq. (\ref{Evolution}) and Eq. (\ref{masterN}) are shown in Fig. \ref{NoReset}(d), and the importance for periodic reinitializing of electron spin is illustrated.

We consider the same case of a silicon-29 and a carbon-13 spin coupled to an NV spin which is presented in Section II except that the NV center is reinitialized to state $|-\rangle_{e}$ every  $t_{re}=20$ $\mu s$. The resonant condition is matched as $\Omega_{rf1}=\Omega_{rf2}=(2\pi)1$ kHz, both of the simulations of exact numerical calculations and effective master equation show perfect state transfer between nuclear spins silicon-29 and carbon-13 $|+_{1}-_{2}\rangle\rightarrow|-_{1}+_{2}\rangle$, [see Fig. \ref{NoReset}(d)]. The state evolutions based on effective master equation are slightly different from the exact simulations. The possible reason is high orders of SW transformation expansions are not included.

In the absence of periodic reinitialization of the electron spin, channel mixing up leads to no coherent evolution of the nuclear spins  [see Fig. \ref{NoReset}(c)], while periodic reinitialization of the NV center to state $|-\rangle_{e}$ provides a priority channel during the operations and coherent evolution between the nuclear spins is possible to extend beyond the life time of the NV spin. High fidelity of the nuclear-nuclear quantum gate in one-step is possible ($>$0.99), which could be significantly higher than the fidelity ($<$0.66) of nuclear-nuclear gates achieved so far with NV centers by using 4 electron-nuclear spin quantum gates (if each fidelity $<$0.90 \cite{Taminiau2014UniversalCA}). Additionally, because of the second-order coupling, the scheme is not sensitive to the detunings from the resonance and Rabi frequency errors of the MW driving of the NV center [see Fig. \ref{NoReset}(e)]. The process fidelity of two nuclear spins is given by the overlap of the states corresponding to the implemented evolution $|\psi\rangle=U^{zz}\otimes\sum_{i,j}|i_1\rangle|j_2\rangle/4$ (i,j=+,-) to the target states.

\section{Quantum gate implementation}
In the Fig. \ref{Operation}, we include all the related operations of the nuclear ZZ gate implementation for experiments. The initialization consists of the electron (nuclear) spin polarization P (M). The electron spin of NV center can be optically initialized and read out by using laser illumination. Here, P is obtained by the optical pumping cycle available for NV centers \cite{Jelezko2002SingleSS,Jelezko2004ObservationOC}, whereas M is based on the techniques developed for the nuclear single-shot measurement \cite{Neumann2010SingleShotRO}, followed by the electron state-dependent fluorescence \cite{Jelezko2002SingleSS,Jelezko2004ObservationOC}. Gate operations includes 3 steps: (i) Polarize the carbon-13 and silicon-29 nuclear spins by using the NV center, which leads to the initial state $|-\rangle_{e}\otimes|+_{1}-_{2}\rangle$ of the system. (ii) Implement the gate operation, which is governed by the effective master equation of Eq. (\ref{masterN}), $U^{ZZ}(t)=\exp(-iH_{N}t)$. In this process, we periodically reset the NV center and we control the heteronuclear spins via RF fields individually. (iii) Use the NV center to read out the states of nuclear spins. Here we suppose that the carbon-13 and silicon-29 spin can be well controlled by a SWAP gate between NV and nuclear spin; i.e., it can be polarized and its position can be detected precisely.

Assuming that the RF field is applied to the carbon-13 spin resonantly and $\delta_2=\gamma_{n2}B+\frac{a_{\parallel2}}{2}-\omega_{rf2}$ is a detuning from the resonance of RF driving to the silicon-29 spin, the effective Hamiltonian is given by
\begin{eqnarray}\label{Detuning}
H'_{N}\approx\bar{\Omega}_{rf2}I^{x'}_{2}+\Omega_{rf1}I^{x}_{1}+pg'_{e}I^{z}_{1}I^{z}_{2},
\end{eqnarray}
in which $\bar{\Omega}_{rf2}=\sqrt{\delta_2^2+\Omega_{rf2}^2}$. The initial state of the system is $|+_1-_2\rangle$, the population ($P_+$) of the initial state $|+_1\rangle$ of carbon-13 spin is given by
\begin{eqnarray}\label{P}
P_+=1-\frac{(pg'_{e}\cos\theta)^{2}\sin^{2}[t
\sqrt{(pg'_{e}\cos\theta)^{2}+\Delta^{2}}/2]}{(pg'_{e}\cos\theta)^{2}+\Delta^{2}},
\end{eqnarray}
$\cos\theta=\frac{\Omega_{rf2}}{\sqrt{\delta_2^2+\Omega_{rf2}^2}}$ and $\Delta=\bar{\Omega}_{rf2}-\Omega_{rf1}$. The frequency of the RF driving to silicon-29 spin is swept to show the detection bandwidth of our method [see Fig. \ref{Robust}]. The perfect population transfer could be possible if $|\bar{\Omega}_{rf2}-\Omega_{rf1}|\ll pg'_{e}\cos\theta$. Namely, the dip position with $\omega_{rf2}=\gamma_{n2}B+\frac{a_{\parallel2}}{2}=(2\pi)5065.5$kHz indicates $\delta_2=0$ and $\Omega_{rf2}=\Omega_{rf1}$. The dip height depends on the longitudinal dipolar coupling with $P_+=\cos^{2}(pg'_{e}T/2)$ with $T$ the total evolution time. Increasing the Rabi frequency of the MW driving induces a smaller effective coupling between the nuclear spins, decreasing the frequency bandwidth [see Fig. \ref{Robust}], which also gives a tunable frequency filter. We investigate the case when there is a third silicon-29 nuclear spin with coupling $a_{\parallel_3}$ to NV center and show the efficiency of the selectivity of our scheme. In Fig. \ref{Robust}(b), we initialize the nuclei in state $|+_{1}-_{2}-_{3}\rangle$. We calculate the infidelity of nuclear ZZ gate between two nuclear spins affected by the third silicon-29 spin. The first carbon-13 and second silicon spins are coupled to an NV center with $[a_{\parallel 1},a_{\parallel 2}]=(2\pi)[11,9]$ kHz. The coupling of the third spin matches $\delta_{3}=(a_{\parallel 2}-a_{\parallel 3})/2>(2\pi)0.5$ kHz gives high fidelity of the gate operation of target nuclear spins ($>0.95$). Thus, our method can address two heteronuclear spins by using the RF fields individually, and implement a near perfect quantum gate at room temperature with high selectivity. 

\section{Sensing application}

\begin{figure}
\center
\includegraphics[width=3.0in]{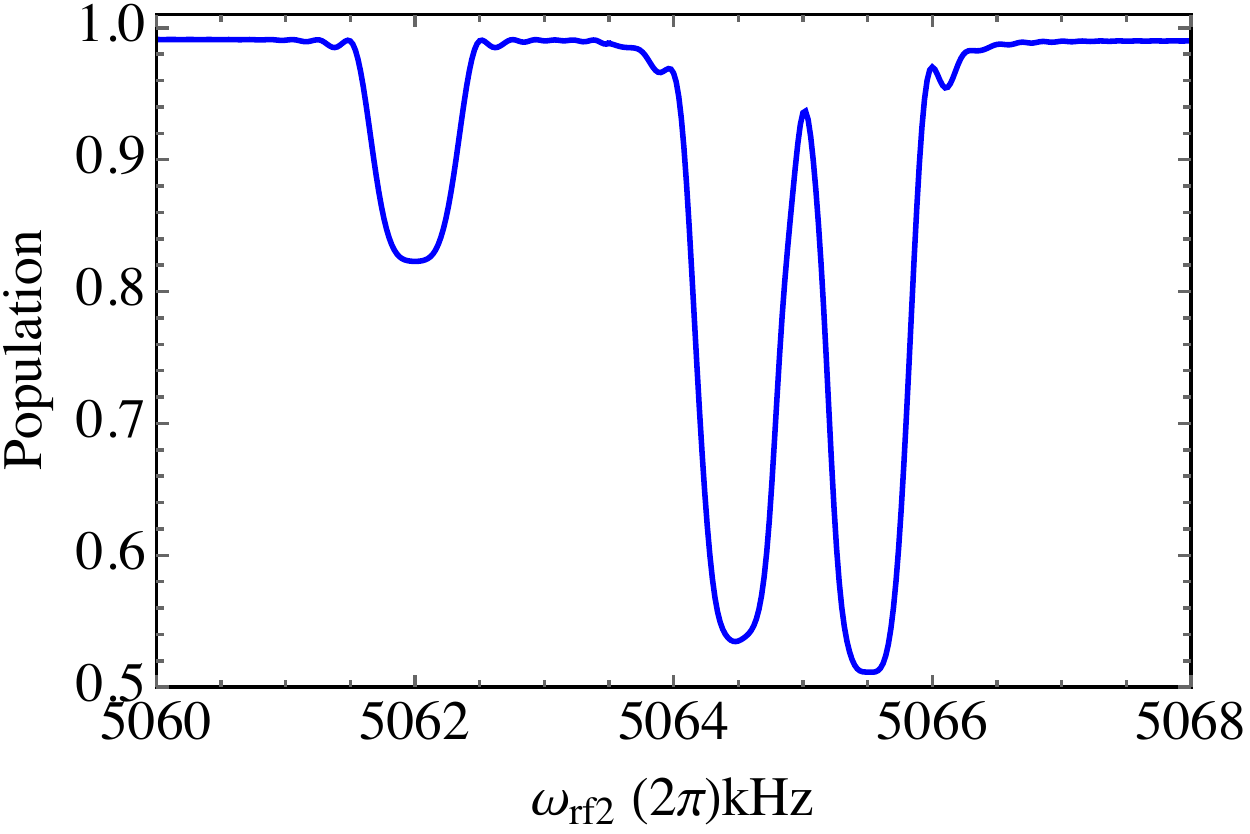}
\caption{Assuming a carbon-13 and three hydrogen-1 spins coupled to the NV center with parallel couplings $[a_{\parallel 1},a_{\parallel 2},a_{\parallel 3},a_{\parallel 4}]=(2\pi)[11,4,9,11]$ kHz. The carbon-13 spin is the sensor with the initial state $|+_{1}\rangle$ and the other three hydrogen-1 are the targets which are in maximum mixed states. The population of the initial state $|+_{1}\rangle$ of the carbon-13 sensor is calculated by exact simulation of Eq. (\ref{Evolution}) with evolution time $T=8$ ms for $\Omega=(2\pi)400$ kHz and the target spins decoherence time $T_2=200$ ms.  }
\label{Third}
\end{figure}

Another important application of our scheme is to detect nuclear spins in another species outside of the diamond. For example, one can use a carbon-13 spin which is well controlled by the NV center to detect hydrogen-1 spins outside of the diamond. Thus carbon-13 spin is initialized in state $|+_{1}\rangle$ and hydrogen-1 spins are in maximally mixed states, one can detect the signal by measuring the probability of that the sensor carbon-13 spin remains in the state $|+_{1}\rangle$. The related operations are shown in Fig. \ref{Operation}, except of the initialization and readout of hydrogen-1 spins. To evaluate the effectiveness of our sensing application, we initialize the carbon-13 spin to $|+_{1}\rangle$ state to detect the other three nuclear (hydrogen-1) spins, see Fig. \ref{Third}. Based on Eq. (\ref{P}), we can adjust the parallel components and control frequency of RF to match the resonance condition for target (hydrogen-1) nuclear spins. Once the resonance condition is achieved, the signal dips mark the presence of the nuclear spins. Similar to the case in the previous scheme  \cite{Chen2017DissipativelySQ}, the frequency resolution is not limited by the NV relaxation $T_{1\rho}$, it requires $|\bar{\Omega}_{rfi}-\Omega_{rf1}|>pg'_{e}\cos\theta$ which is limited by the target spin decoherence time $T_2=200$ ms, the sensitivity per unit time of our scheme is proportional to $pg'_{e}/\sqrt{1/\Gamma_N}\sim(a_{\parallel i}/4)/\sqrt{T_{1\rho}}.$

\section{Conclusion}
Before summarizing, we would like to compare our new method with the previous scheme \cite{Chen2017DissipativelySQ}. Both of schemes realize the effective coupling of target nuclear spins by periodical resetting of electron spin, and the impact of NV decoherence and relaxation processes on these nuclei are suppressed. The difference mainly comes from that we employ two weak RF fields to control the heteronuclear spins individually.  In the previous approach, nuclear spins in the same species are considered, coarse tuning of the direction of the magnetic field in advance is necessary for matching the resonant condition. Our method allows for effective interaction of nuclear spins in different species which loosen the requirement of the stringent resonant condition of the previous scheme. The signal accumulation and resonant condition depend on longitudinal dipolar coupling component $a_{\parallel_i}$, which leads to no information of the transverse coupling component $a_{\perp_i}$ to be detected in this new scheme. Therefore, less information is detected when one uses this new method for sensing nuclear spins near the NV center. But it also bring advantage that the quantum gate implementation is not limited by the transverse coupling component. 

In this work we presented an extension of dissipatively stabilized NV center to a case, in which one achieves a tunable second-order effective coupling between distant nuclear spins in different species. We employ RF fields and MW driving to match the resonant condition, and control electron and nuclear spins individually. The coupling is mediated by a dissipatively decoupled electron spin of a NV center, when the impact of NV decoherence and relaxation processes on these nuclei are suppressed. Thus the effective indirect interaction enables selectively initialization and coherent control of nuclear spins as well as analysis of complex spin structures at ambient condition.

\emph{Acknowledgements ---} The authors thank Natural Science Foundation of Hunan Province, China (2019JJ10002), Hunan Provincial Hundred People Plan (2019), Huxiang High-level Talent Gathering Project (2019RS1043).


\begin{thebibliography}{99}
\bibitem{Taminiau2014UniversalCA} T. Taminiau, J. Cramer, T. van der Sar, V. Dobrovitski, and R. Hanson, Nature nanotechnology \textbf{9 3}, 171 (2014).
\bibitem{2018Pulse} J. Zhang, S. S. Hegde, and D. Suter, Physical Review A \textbf{98}, 042302 (2018).
\bibitem{Waldherr2014QuantumEC} G. Waldherr, Y. Wang, S. Zaiser, M. Jamali, T. Schulte-Herbr\"{u}ggen, H. Abe, T. Ohshima, J. Isoya, J. F. Du, P. Neumann, et al., Nature \textbf{506}, 204 (2014).
\bibitem{2011High} L. Robledo, L. Childress, H. Bernien, B. Hensen, P. Alke-made, and R. Hanson, Nature \textbf{477}, 574 (2011).
\bibitem{Taminiau2012DetectionAC} T. Taminiau, J. Wagenaar, T. van der Sar, F. Jelezko, V. Dobrovitski, and R. Hanson, Physical review letters \textbf{109 13}, 137602 (2012).
\bibitem{Kolkowitz2012SensingDN} S. Kolkowitz, Q. Unterreithmeier, S. Bennett, and M. Lukin, Physical review letters \textbf{109 13}, 137601 (2012).
\bibitem{Zhao2012SensingSR} N. Zhao, J. Honert, B. Schmid, M. Klas, J. Isoya, M. Markham, D. Twitchen, F. Jelezko, R.-B. Liu, H. Fedder, et al., Nature nanotechnology \textbf{7 10}, 657 (2012).
\bibitem{Ermakova2013DetectionOA} A. Ermakova, G. Pramanik, J.-M. Cai, G. Algara-Siller, U. Kaiser, T. Weil, Y.-K. Tzeng, H. C. Chang, L. McGuinness, M. Plenio, et al., Nano letters \textbf{13 7}, 3305 (2013).
\bibitem{Mamin2013NanoscaleNM} H. J. Mamin, M. Kim, M. Sherwood, C. Rettner, K. Ohno, D. Awschalom, and D. Rugar, Science \textbf{339}, 557 (2013).
\bibitem{Zaiser2016EnhancingQS} S. Zaiser, T. Rendler, I. Jakobi, T. Wolf, S.-Y. Lee, S.Wagner, V. Bergholm, T. Schulte-Herbr\"{u}ggen, P. Neumann, and J. Wrachtrup, Nature Communications \textbf{7} (2016).
\bibitem{Staudacher2013NuclearMR} T. Staudacher, F. Shi, S. Pezzagna, J. Meijer, J. Du, C. A. Meriles, F. Reinhard, and J. Wrachtrup, Science \textbf{339}, 561 (2013).
\bibitem{Wu2016DiamondQD} Y. Wu, F. Jelezko, M. Plenio, and T. Weil, Angewandte Chemie \textbf{55 23}, 6586 (2016).
\bibitem{Doherty2013TheNC} M. Doherty, N. Manson, P. Delaney, F. Jelezko, J. Wrachtrup, and L. Hollenberg, Physics Reports \textbf{528}, 1 (2013).
\bibitem{2016Optomechanical} D. A. Golter, T. Oo, M. Amezcua, K. A. Stewart, and H. Wang, Physical review letters \textbf{1116}, 143602 (2016).
\bibitem{2014Optically}D. A. Golter and H. Wang, Physical review letters \textbf{112}, 116403 (2014).
\bibitem{Casanova2016NoiseResilientQC} J. Casanova, Z.-Y. Wang, and M. Plenio, Physical review letters \textbf{117 13}, 130502 (2016).
\bibitem{Bermdez2011ElectronmediatedNI} A. Berm\'{u}dez, F. Jelezko, M. Plenio, and A. Retzker, Physical review letters \textbf{107 15}, 150503 (2011).
\bibitem{zimmermann2020selective} J. Zimmermann, P. London, Y. Yirmiyahu, F. Jelezko, A. Blank, and D. Gershoni, Physical Review B \textbf{102}, 245408 (2020).
\bibitem{rong2014implementation} X. Rong, J. Geng, Z. Wang, Q. Zhang, C. Ju, F. Shi, C.K. Duan, and J. Du, Physical review letters \textbf{112}, 050503 (2014).
\bibitem{hegde2020efficient} S. S. Hegde, J. Zhang, and D. Suter, Physical review letters \textbf{124}, 220501 (2020).
\bibitem{abobeih2019atomic}M. Abobeih, J. Randall, C. Bradley, H. Bartling, M. Bakker, M. Degen, M. Markham, D. Twitchen, and T. Taminiau, Nature \textbf{576}, 411 (2019).
\bibitem{Bradley2019ATS}C. Bradley, J. Randall, M. Abobeih, R. Berrevoets, M. Degen, M. Bakker, M. Markham, D. Twitchen, and T. Taminiau, Physical Review X \textbf{9}, 031045 (2019).
\bibitem{Wang2017DelayedEE} Z.-Y. Wang, J. Casanova, and M. Plenio, Nature Communications \textbf{8} (2017).
\bibitem{Tratzmiller2021ParallelSN}B. Tratzmiller, J. Haase, Z.-Y. Wang, and M. Plenio, Physical Review A \textbf{103}, 012607 (2021).
\bibitem{Casanova2017ArbitraryNG} J. Casanova, Z.-Y. Wang, and M. Plenio, Physical Review A \textbf{96}, 032314 (2017).
\bibitem{degen2021entanglement} M. Degen, S. Loenen, H. Bartling, C. Bradley, A. Meinsma, M. Markham, D. Twitchen, and T. Taminiau, Nature Communications \textbf{12}, 1 (2021).
\bibitem{Bian2017UniversalQC} J. Bian, M. Jiang, J. Cui, X. Liu, B. Chen, Y. Ji, B. Zhang, J. Blanchard, X. Peng, and J. Du, Physical Review A \textbf{95}, 052342 (2017).
\bibitem{Chen2017DissipativelySQ} Q. Chen, I. Schwarz, and M. Plenio, Physical review letters \textbf{119 1}, 010801 (2017).
\bibitem{Kessler2012GeneralizedSF} E. Kessler, Physical Review A \textbf{86}, 012126 (2012).
\bibitem{Bravyi2011SchriefferWolffTF} S. Bravyi, D. DiVincenzo, and D. Loss, Annals of Physics \textbf{326}, 2793 (2011).
\bibitem{Reiter2012EffectiveOF} F. Reiter and A. S{\o}rensen, Physical Review A \textbf{85}, 032111 (2012).
\bibitem{Jelezko2002SingleSS} F. Jelezko, I. Popa, A. Gruber, C. Tietz, J. Wrachtrup, A. Nizovtsev, and S. Kilin, Applied Physics Letters \textbf{81}, 2160 (2002).
\bibitem{Jelezko2004ObservationOC} F. Jelezko, T. Gaebel, I. Popa, A. Gruber, and J. Wrachtrup, Physical review letters \textbf{92 7}, 076401 (2004).
\bibitem{Neumann2010SingleShotRO} P. Neumann, J. Beck, M. Steiner, F. Rempp, H. Fedder, P. Hemmer, J. Wrachtrup, and F. Jelezko, Science \textbf{329}, 542 (2010).
\bibitem{Laraoui2013HighresolutionCS} A. Laraoui, F. Dolde, C. Burk, F. Reinhard, J. Wrachtrup, and C. A. Meriles, Nature Communications \textbf{4}, 1651 (2013).
\bibitem{Pfender2017NonvolatileNS} M. Pfender, N. Aslam, H. Sumiya, S. Onoda, P. Neumann, J. Isoya, C. A. Meriles, and J. Wrachtrup, Nature Communications \textbf{8} (2017).
\bibitem{Cai2013DiamondbasedSM} J. Cai, F. Jelezko, M. Plenio, and A. Retzker, New Journal of Physics \textbf{15}, 013020 (2013).

\end{thebibliography}
\end{document}